\documentclass{ws-rv9x6}
\usepackage{subfigure}     
\usepackage{ws-rv-van}     
\makeindex
\newcommand{\be}{\begin{equation}}
\newcommand{\ee}{\end{equation}}
\newcommand{\bea}{\begin{eqnarray}}
\newcommand{\eea}{\end{eqnarray}}
\begin{document}

\chapter[Relativistic QRPA and
quasiparticle-vibration coupling]{Microscopic description of nuclear
vibrations: \\ Relativistic QRPA and its extensions \\ with
quasiparticle-vibration coupling }\label{ra_ch1}

\author[Elena Litvinova and Victor Tselyaev]{Elena Litvinova
}

\address{National Superconducting Cyclotron Laboratory, Michigan State
University, East Lansing, MI 48824-1321, USA
\\
Litvinova@nscl.msu.edu
}

\author[Elena Litvinova and Victor Tselyaev]{Victor Tselyaev}
\address{Nuclear Physics Department, St. Petersburg State
University, \\198504 St. Petersburg, Russia\\
tselyaev@nuclpc1.phys.spbu.ru
}
\begin{abstract}
The recent extensions of the covariant energy density functional
theory with the quasiparticle-vibration coupling (QVC) are reviewed.
Formulation of the Quasiparticle Random Phase Approximation (QRPA)
in the relativistic framework is discussed. Self-consistent
extensions of the relativistic QRPA imply the QVC which is
implemented in two-body propagators in the nuclear medium. This
provides fragmentation of the QRPA states describing the damping of
the vibrational motion.
\end{abstract}
\body
\section{Introduction}
Shortly after the appearance of the Bardeen-Cooper-Schrieffer (BCS)
theory of superconductivity \cite{BCS.57}, Bohr, Mottelson and Pines
have noticed that atomic nuclei exhibit properties similar to a
superconducting metal \cite{BMP.58}. An energy gap between the
ground state and the first intrinsic excitation is found to be a
common feature of Fermi-systems with an interaction acting between
particles with equal and opposite momenta. Such {\it pairing
correlations} in nuclei are responsible for the reduction of nuclear
moments of inertia, compared to the case of rigid rotation, and
intimately connected to odd-even mass differences, low-lying
vibrational states, nuclear shapes and level densities \cite{RS.80}.
Over the decades, starting from the works \cite{BMP.58,S.59,B.59},
the BCS and the more general Bogoliubov's concept \cite{Bo.58} are
widely used for the description of ground state properties of
open-shell nuclei. For nuclear excited states, the straightforward
generalization of the random phase approximation (RPA) \cite{BP.53},
the quasiparticle RPA (QRPA)\cite{Bo.59,Ba.60,So.71} including
pairing correlations has become a standard approach.

Impressive progress of experimental low-energy nuclear physics such
as synthesis of many exotic nuclei \cite{GG.08} and discovering new
nuclear structure phenomena \cite{A.08} insistently calls for
conceptually new theoretical methods. High-precision description of
nuclear properties still remains a challenge for contemporary
theoretical physics. One of the most promising strategies for
medium-mass and heavy nuclei is the construction of a "universal"
nuclear energy density functional supplemented by various many-body
correlations. A delicate interplay of different kinds of
correlations is responsible for binding loosely-bound systems, decay
properties and for low-energy spectra.

The first fully self-consistent QRPA \cite{PRN.03} has been
developed on the base of the covariant energy density functional
(CEDF) \cite{VALR.05} with pairing correlations described by the
pairing part of the finite-range Gogny interaction. The great
success of the RQRPA
in applications to various nuclear structure phenomena has
emphasized the importance of the self-consistency between the mean
field and the effective interaction. Our recent attempts to include
correlations beyond the CEDF and the RQRPA use the relativistic
framework \cite{Rin.96,VALR.05} in combination with advancements of
the Landau - Migdal theory for Fermi liquids in parameter-free field
theory techniques \cite{LRT.08,LRT.10,Lit.12}. Couplings of
single-particle and collective degrees of freedom are included on
equal footing with the pairing correlations in a fully
self-consistent way. In this Chapter we give a brief review of these
developments.
\section{Covariant energy density functional theory with pairing correlations}
In contrast to Hartree or Hartree-Fock theory, where the building
blocks of excitations (the quasiparticles in the sense of Landau)
are either nucleons in levels above the Fermi surface (particles) or
missing nucleons in levels below the Fermi surface (holes),
quasiparticles in the sense of Bogoliubov are described by a
combination of creation and annihilation operators. This fact can be
expressed, following Nambu and Gor'kov \cite{AGD.63}, by introducing
the following two-component operator, which is a generalization of
the usual particle annihilation operator:
\begin{equation}
\Psi(1)=\left(
\begin{array}
[c]{c}%
a(1)\\
a^{\dagger}(1)
\end{array}
\right)  . \label{psi}%
\end{equation}
Here $a(1)=e^{iHt_{1}}a_{k_{1}}e^{-iHt_{1}}$ is a nucleon
annihilation operator in the Heisenberg picture and the quantum
numbers $k_{1}$ represent an arbitrary basis, $1=\{k_{1},t_{1}\}$.
In order to keep the notation simple we use in the following
$1=\{\mbox{\boldmath $r$}_{1},t_{1}\}$ and omit spin and isospin
indices.

Let us introduce the chronologically ordered product of the operator
$\Psi(1)$ in Eq. (\ref{psi}) and its Hermitian conjugated operator
$\Psi^{\dagger}(2)$, averaged over the ground state
$|\Phi_{0}\rangle$ of a nucleus. This tensor of rank 2
\begin{equation}
G(1,2)=-i\langle\Phi_{0}|T\Psi(1)\Psi^{\dagger}(2)|\Phi_{0}\rangle
\ \ \label{fg}%
\end{equation}
is the generalized Green's function which can be expressed through a
2$\times $2 matrix:
\begin{align}
G(1,2)  &  =-i\theta(t_{1}-t_{2})\langle\Phi_{0}|\left(
\begin{array}
[c]{cc}%
a(1)a^{\dagger}(2) & a(1)a(2)\\
a^{\dagger}(1)a^{\dagger}(2) & a^{\dagger}(1)a(2)
\end{array}
\right)  |\Phi_{0}\rangle\nonumber\\
&  +i\theta(t_{2}-t_{1})\langle\Phi_{0}|\left(
\begin{array}
[c]{cc}%
a^{\dagger}(2)a(1) & a(2)a(1)\\
a^{\dagger}(2)a^{\dagger}(1) & a(2)a^{\dagger}(1)
\end{array}
\right)  |\Phi_{0}\rangle. \label{gmat}%
\end{align}
Therefore, the generalized density matrix is obtained as a limit
\begin{equation}
\mathcal{R}(\mbox{\boldmath $r$}_{1},\mbox{\boldmath
$r$}_{2},t_{1})=-i\lim\limits_{t_{2}\rightarrow t_{1}+0}G(1,2) \label{limg}%
\end{equation}
from the second term of Eq. (\ref{gmat}), and, in the notation of
Valatin \cite{Val.61}, it can be expressed as a matrix of doubled
dimension containing as components the normal density ${{\rho}}$ and
the abnormal density ${{\varkappa}}$, the so called pairing tensor:
\begin{equation}
\mathcal{R}(\mbox{\boldmath $r$}_{1},\mbox{\boldmath
$r$}_{2},t)=\left(
\begin{array}
[c]{cc}%
\rho(\mbox{\boldmath $r$}_{1},\mbox{\boldmath $r$}_{2},t) & \ \ \ \
\varkappa(\mbox{\boldmath $r$}_{1},\mbox{\boldmath
$r$}_{2},t)\\
-\varkappa^{\ast}(\mbox{\boldmath $r$}_{1},\mbox{\boldmath
$r$}_{2},t) & \ \ \ \ \delta(\mbox{\boldmath
$r$}_{1}-\mbox{\boldmath $r$}_{2})-\rho^{\ast}(\mbox{\boldmath
$r$}_{1},\mbox{\boldmath $r$}_{2},t)
\end{array}
\right)  . \label{rvalatin}%
\end{equation}
These densities play a key role in the description of a superfluid
many-body system.

In CEDF theory for normal systems the ground state of a nucleus is a
Slater determinant describing nucleons, which move independently in
meson fields $\phi_{m}$ characterized by their quantum numbers for
spin, parity and isospin. In the present investigation we use the
concept of the conventional relativistic mean field (RMF) theory and
include the $\sigma$, $\omega$, $\rho$-meson fields and the
electromagnetic field as the minimal set of fields providing a
rather good quantitative description of bulk and single-particle
properties~in the nucleus \cite{Wal.74,SW.86,Rin.96,VALR.05}. This
means that the index $m$ runs over the different types of fields
$m=\{\sigma,\omega,\rho,A\}$. The summation over $m$ implies in
particular scalar products in Minkowski space for the vector fields
and in isospace for the $\rho$-field.

The total energy depends in the case without pairing correlations on
the normal density matrix ${\rho}$\ and the various fields
$\phi_{m}$:
\bea E_{RMF}[{{\rho}},\phi]&=&\text{Tr}[({\mbox{\boldmath
$\alpha$}}\mathbf{p}+\beta m){{\rho}}] + \nonumber\\ %
&+& \sum\limits_{m}\Bigl\{\text{Tr}[(\beta\Gamma_{m}\phi_{m}){{\rho}%
}]\pm\int\Bigl[\frac{1}{2}(\mbox{\boldmath $\nabla$}\phi_{m})^{2}+U_{m}%
(\phi)\Bigr]d^{3}r\Bigr \}. \label{ERMF}%
\eea
Here we have neglected retardation effects, i.e. time-derivatives of
the fields $\phi_{m}$. The plus sign in Eq.~(\ref{ERMF}) holds for
scalar fields and the minus sign for vector fields. The trace
operation implies a sum over Dirac indices and an integral in
coordinate space. ${\mbox{\boldmath $\alpha$}}$ and $\beta$ are
Dirac matrices and the vertices $\Gamma_{m}$ are given by
\begin{equation}
\Gamma_{\sigma}=g_{\sigma},\ \ \ \
\Gamma_{\omega}^{\mu}=g_{\omega}\gamma ^{\mu},\ \ \ \
{\vec{\Gamma}}_{\rho}^{\ \mu}=g_{\rho}{\vec{\tau}}\gamma^{\mu
},\ \ \ \ \Gamma_{e}^{\mu}=e\frac{(1-\tau_{3})}{2}\gamma^{\mu} \label{gammas}%
\end{equation}
with the corresponding coupling constants $g_{m}$ for the various
meson fields and for the electromagnetic field.

$\bigskip$The quantities $U_{m}(\phi)$ are, in the case of a linear
meson couplings, given by the term $
U_{m}(\phi)=\frac{1}{2}m_{m}^{2}\phi_{m}^{2}$ containing the meson
masses $m_{m}$. For non-linear meson couplings, as for instance for
the $\sigma$-meson in the parameter set NL3 we have, as proposed in
Ref.~\cite{BB.77}: $
U(\sigma)=\frac{1}{2}m_{\sigma}^{2}\sigma^{2}+\frac{g_{2}}{3}\sigma^{3}%
+\frac{g_{3}}{4}\sigma^{4}\;. $ with two additional coupling
constants $g_{2}$ and $g_{3}$.

In the superfluid CEDF theory the energy is, in general, a
functional of the Valatin density $\mathcal{{R}}$ and the fields
$\phi_{m}$.
In the present applications we consider a density functional of the
relativistic Hartree-Bogoliubov (RHB) form:
\begin{equation}
E_{RHB}[{{\rho}},{{\varkappa}},{{\varkappa}}^{\ast},\phi]=E_{RMF}[{{\rho}%
},\phi]+E_{pair}[{{\varkappa},{\varkappa}}^{\ast}] \label{ERHB}%
\end{equation}
where the pairing energy is expressed by an effective interaction ${\tilde{V}%
}^{pp}$ in the $pp$-channel:
\begin{equation}
E_{pair}[{{\varkappa},{\varkappa}}^{\ast}]=\frac{1}{4}Tr[{
{\varkappa}}^{\ast
}{\tilde{V}}^{pp}{{\varkappa}}], \label{Epair}%
\end{equation}
assuming no explicit dependence of the pairing part on the nucleonic
density and meson fields. Generally, the form of $\tilde{V}^{pp}$ is
restricted only by the conditions of the relativistic invariance of
$E_{pair}$ with respect to the transformations of the abnormal
densities~\cite{CG.99}. As discussed in \cite{VALR.05}, in the early
applications the same effective Lagrangian was used in both $ph$ and
$pp$ channels, however, such approaches produced too large pairing
gaps, as compared to empirical ones. The reason is the unphysical
behavior of such forces at large momenta. %
In this section, we consider the general form of $\tilde{V}^{pp}$ as
a non-local function in coordinate representation. In the
applications we use for $\tilde {V}^{pp}$ a simple monopole-monopole
interaction \cite{LRT.08}.

The classical variational principle applied to the energy functional
\ref{ERHB} leads to the relativistic Hartree-Bogoliubov equations:
\cite{KuR.91}
\begin{equation}
\mathcal{H}_{RHB}|\psi_{k}^{\eta}\rangle=\eta E_{k}|\psi_{k}^{\eta}%
\rangle,\ \ \ \ \eta=\pm1 \label{hb}%
\end{equation}
with the RHB Hamiltonian
\begin{equation}
{{\mathcal{H}}}_{RHB}=2\frac{\delta
E_{RHB}}{\delta\mathcal{R}}=\left(
\begin{array}
[c]{cc}%
h^{\mathcal{D}}-m-\lambda & \Delta\\
-\Delta^{\ast} & -h^{\mathcal{D}\ast}+m+\lambda
\end{array}
\right)  , \label{HRHB}%
\end{equation}
where $\lambda$ is the chemical potential (counted from the
continuum limit), and $h^{\mathcal{D}}$ is the single-nucleon Dirac
Hamiltonian
\begin{equation}
h^{\mathcal{D}}={\mbox{\boldmath $\alpha$}}\mathbf{p}+\beta(m+{\tilde{\Sigma}%
}), \ \ \ \ \ {\tilde{\Sigma}}(\mbox{\boldmath
$r$})=\sum\limits_{m}\Gamma_{m}\phi
_{m}(\mbox{\boldmath $r$}). \label{hd}%
\end{equation}
The pairing field $\Delta$ reads in this case:
\begin{equation}
\Delta(\mbox{\boldmath $r$},\mbox{\boldmath
$r$}^{\prime})=\frac{1}{2}\int d{\mbox{\boldmath
$r$}}^{\prime\prime}d{\mbox{\boldmath $r$}}^{\prime\prime\prime}{\tilde{V}%
}^{pp}(\mbox{\boldmath $r$},\mbox{\boldmath $r$}^{\prime}%
,\mbox{\boldmath $r$}^{\prime\prime},\mbox{\boldmath
$r$}^{\prime\prime\prime})\varkappa(\mbox{\boldmath
$r$}^{\prime\prime},\mbox{\boldmath $r$}^{\prime\prime\prime}), \label{del}%
\end{equation}
and the generalized density matrix
\begin{equation}
\mathcal{R}(\mbox{\boldmath $r$},\mbox{\boldmath $r$}^{\prime})=\sum
\limits_{k}|\psi_{k}^{-}(\mbox{\boldmath $r$})\rangle\langle\psi_{k}%
^{-}(\mbox{\boldmath $r$}^{\prime})| \label{gdm}
\end{equation}
is composed from the 8-dimensional Bogoliubov-Dirac spinors of the
following form:
\begin{equation}
|\psi_{k}^{+}(\mbox{\boldmath $r$})\rangle=\left(
\begin{array}
[c]{c}%
U_{k}(\mbox{\boldmath $r$})\\
V_{k}(\mbox{\boldmath $r$})
\end{array}
\right)  ,\ \ \ \ |\psi_{k}^{-}(\mbox{\boldmath $r$})\rangle=\left(
\begin{array}
[c]{c}%
V_{k}^{\ast}(\mbox{\boldmath $r$})\\
U_{k}^{\ast}(\mbox{\boldmath $r$})
\end{array}
\right)  . \label{dbasis}%
\end{equation}
In Eq. (\ref{gdm}), the summation is performed only over the states
having large upper components of the Dirac spinors.
This restriction corresponds to the so-called \textit{no-sea
approximation}~\cite{SR.02}.

The behavior of the meson and Coulomb fields is derived from the
energy functional (\ref{ERHB}) by variation with respect to the
fields $\phi_{m}$. We obtain Klein-Gordon equations. In the static
case they have the form:
\begin{equation}
-\Delta\phi_{m}(\mbox{\boldmath $r$})+U^{\prime}(\phi_{m}%
(\mbox{\boldmath $r$}))=\mp\sum\limits_{k}V_{k}^{\intercal}%
(\mbox{\boldmath $r$})\beta\Gamma_{m}V_{k}^{\ast}(\mbox{\boldmath
$r$}).
\label{kg}%
\end{equation}
Eq. (\ref{kg}) determines the potentials entering the single-nucleon
Dirac Hamiltonian (\ref{hd}) and is solved self-consistently
together with Eq. (\ref{hb}). The system of Eqs. (\ref{hb}) and
(\ref{kg}) determine the ground state of an open-shell nucleus in
the RHB approach. In the following, however, we use the Hartree-BCS
approximation, where the Dirac hamiltonian $h^{\cal D}$ (\ref{hd})
and the normal nucleon density $\rho$ are diagonal. In this
approximation the spinors (\ref{dbasis}) are expressed through
eigenvectors of the operator $h^{\cal D}$.
Below we call this basis Dirac-Hartree-BCS (DHBCS) basis.
%
\section{Relativistic QRPA}\label{rqrpa}
Spectra of nuclear excitations are very important for an
understanding of the nuclear structure. Apart from particle-hole or
few-quasiparticle excitations there are also rotational and
vibrational states involving coherent motion of many nucleons. In
spherical nuclei collective vibrations like giant resonances
dominate in nuclear spectra \cite{HW.01}. They are characterized by
high values of electromagnetic transition probabilities and show up
in spectra of various nuclei over the entire nuclear chart
\cite{RS.80}. The random phase approximation, first proposed in Ref.
\cite{BP.53} to describe collective excitations in degenerate
electron gas, is widely used for various kinds of correlated Fermi
systems including atomic nuclei. The Quasiparticle RPA for
superfluid systems has been constructed in a complete analogy to the
normal case \cite{Bo.59,Ba.60,So.71}. The effective field equations
of the Theory of Finite Fermi Systems \cite{Mig.67} developed as an
extension of Landau's theory for Fermi liquid are, in fact, the QRPA
equations.

The derivation of the relativistic QRPA (RQRPA) equations is a
straightforward generalization of the relativistic RPA (RRPA)
\cite{RMGVWC.01} formulated in the doubled space (\ref{dbasis}) of
Bogoliubov quasiparticles. Both RRPA and RQRPA equations are
obtained as a small-amplitude limit of the time-dependent RMF model.
In Ref. \cite{PRN.03} the RQRPA equations are formulated and solved
in the canonical basis of the RHB model.

The key quantity describing an oscillating nuclear system is
transition density $\mathcal{R}_{\mu}$ defined by the harmonic time
dependence of the generalized density matrix (\ref{rvalatin}):
\begin{equation}
\mathcal{R}(t)=\mathcal{R}_{0}+%
{\displaystyle\sum\limits_{\mu}}
(\mathcal{R}_{\mu}e^{i\Omega_{\mu}t}+\text{h.c.)}. \label{rhomu}%
\end{equation}

The general equation of motion for $\mathcal R(t)$
\be i\partial_t{\mathcal R} = [{\cal H}_{RHB}(\cal R),{\cal R}] \ee
and the condition ${\cal R}^2(t) = {\cal R}(t)$ lead in the
small-amplitude limit to the QRPA equation which in the DHBCS basis
has the form::
\begin{equation}
\mathcal{R}_{\mu;k_{1}k_{2}}^{\eta}={\tilde{R}}_{k_{1}k_{2}}^{(0)\eta}%
(\Omega_{\mu})\sum\limits_{k_{3}k_{4}}\sum\limits_{\eta^{\prime}}\tilde
{V}_{k_{1}k_{4},k_{2}k_{3}}^{\eta\eta^{\prime}}\mathcal{R}_{\mu;k_{3}k_{4}%
}^{\eta^{\prime}}, \label{qrpa}%
\end{equation}
where we have introduced the static effective interaction between
quasiparticles $\tilde V$. It is obtained as a functional derivative
of the RMF self-energy ${{\tilde{\Sigma}}}$ with respect to the
relativistic generalized density matrix ${\mathcal{R}}$:
\begin{equation}
{\tilde{V}}_{k_{1}k_{4},k_{2}k_{3}}^{\eta_{1}\eta_{4},\eta_{2}\eta_{3}}%
=\frac{\ \delta{\tilde{\Sigma}}_{k_{4}k_{3}}^{\eta_{4}\eta_{3}}\ }%
{\ \delta\mathcal{R}_{k_{2}k_{1}}^{\eta_{2}\eta_{1}}\ }.
\label{static-interaction}%
\end{equation}
In Eq. (\ref{qrpa}) we denote:
$\mathcal{R}_{\mu;k_{1}k_{2}}^{\eta}~=~\mathcal{R}_{\mu;k_{1}k_{2}}^{\eta,-\eta
},\text{\ \ }{\tilde{R}}_{k_{1}k_{2}}^{(0)\eta}(\omega
)~=~{\tilde{R}}_{k_{1}k_{2}}^{(0)\eta,-\eta}(\omega),\text{}$ and
$\tilde{V}_{k_{1}k_{4},k_{2}k_{3}}^{\eta\eta^{\prime}}~=~\tilde{V}_{k_{1}%
k_{4},k_{2}k_{3}}^{\eta,-\eta^{\prime},-\eta,\eta^{\prime}}$. This
means that we cut out certain components of the tensors in the
quasiparticle space. The quantity ${\tilde{R}}$ is the propagator of
two-quasiparticles in the mean-field, or the mean-field response
function which is a convolution of two single-quasiparticle
mean-field Green's functions (see Eq. (\ref{mfgft}) below):
\begin{equation}
{\tilde{R}}_{k_{1}k_{2}}^{(0)\eta}(\omega)=\frac{1}{\eta\omega-E_{k_{1}%
}-E_{k_{2}}},
\end{equation}
where $E_{k_i}$ are the energies of the Bogoliubov quasiparticles.
%

\section{Beyond RMF: Quasiparticle-vibration coupling model for the nucleon self-energy}

The single-quasiparticle equation of motion (\ref{hb}) determines
the behavior of a nucleon with a static self-energy $\tilde
{\Sigma}$ (\ref{hd}). To include dynamical correlations, i.e. a more
realistic time dependence in the self-energy, one has to extend the
energy functional by appropriate terms. In the present work we use
for this purpose the successful but relatively simple
quasiparticle-vibration coupling (QVC) model introduced in
Refs.~\cite{BM.75,AGD.63}. Following the general logic of this
model, we consider the total single-nucleon self-energy for the
Green's function defined in Eq. (\ref{fg}) as a sum of the RHB
self-energy and an energy-dependent non-local term in the doubled
space:
\begin{equation}
{{\Sigma}}(\mbox{\boldmath $r$},\mbox{\boldmath
$r$}^{\prime};\varepsilon )={{\tilde{\Sigma}}}(\mbox{\boldmath
$r$},\mbox{\boldmath $r$}^{\prime
})+{{\Sigma}^{(e)}}(\mbox{\boldmath $r$},\mbox{\boldmath
$r$}^{\prime
};\varepsilon) \label{Sigma}%
\end{equation}
with
\begin{equation}
{{\tilde{\Sigma}}}(\mbox{\boldmath $r$},\mbox{\boldmath
$r$}^{\prime})=\left(
\begin{array}
[c]{cc}%
\beta{\tilde{\Sigma}}(\mbox{\boldmath $r$})\delta (\mbox{\boldmath
$r$}-\mbox{\boldmath $r$}^{\prime}) & \Delta
(\mbox{\boldmath $r$},\mbox{\boldmath $r$}^{\prime})\\
-\Delta^{\ast}(\mbox{\boldmath $r$},\mbox{\boldmath $r$}^{\prime}) &
-\beta{\tilde{\Sigma}}^{\ast}(\mbox{\boldmath $r$})\delta
(\mbox{\boldmath $r$}-\mbox{\boldmath $r$}^{\prime})
\end{array}
\right)  .
\end{equation}
Here and in the following a tilde sign is used to express the static
character of a quantity, i.e. the fact that it does not depend on
the energy, and the upper index $e$ indicates the energy dependence.
The energy dependence of the operator ${{\Sigma}^{(e)}}%
(\mbox{\boldmath $r$},\mbox{\boldmath $r$}^{\prime};\varepsilon)$
is determined by the QVC model.
In the DHBCS basis its matrix elements are given by
\cite{LR.06,L.12}:
\begin{equation}
\Sigma_{k_{1}k_{2}}^{(e)\eta_{1}\eta_{2}}(\varepsilon)=\sum\limits_{\eta=\pm
1}\sum\limits_{\eta_{\mu}=\pm1}\sum\limits_{k,\mu}\frac{\delta_{\eta_{\mu
},\eta}{\gamma}_{\mu;k_{1}k}^{\eta_{\mu};\eta_{1}\eta}\ {\gamma}%
_{\mu;k_{2}k}^{\eta_{\mu};\eta_{2}\eta\ast}}{\varepsilon-\eta
E_{k}-\eta_{\mu
}(\Omega_{\mu}-i\delta)},\ \ \ \ \delta\rightarrow+0. \label{sgephon}%
\end{equation}
The index $k$ formally runs over all single-quasiparticle states
including antiparticle states with negative energies.
In practical calculations, it is assumed that there are no pairing
correlations in the Dirac sea \cite{SR.02} and the orbits with
negative energies are treated in the no-sea approximation, although
the numerical contribution of the diagrams with intermediate states
$k$ with negative energies is very small due to the large energy
denominators in the corresponding terms of the self-energy
(\ref{sgephon}) \cite{LR.06}. The index $\mu$ in Eq.~(\ref{sgephon})
labels the set of phonons taken into account. $\Omega_{\mu}$ are
their frequencies and $\eta_{\mu}=\pm1$ labels forward and backward
going components in Eq. (\ref{sgephon}). The vertices
${\gamma}_{\mu;k_{1}k_{2}}^{\eta_{\mu};\eta _{1}\eta_{2}}$ determine
the coupling of the quasiparticles to the collective vibrational
state (phonon) $\mu$:
\begin{equation}
{\gamma}_{\mu;k_{1}k_{2}}^{\eta_{\mu};\eta_{1}\eta_{2}}=\delta_{\eta_{\mu}%
,+1}{\gamma}_{\mu;k_{1}k_{2}}^{\eta_{1}\eta_{2}}+\delta_{\eta_{\mu},-1}%
{\gamma}_{\mu;k_{2}k_{1}}^{\eta_{2}\eta_{1}\ast}.
\end{equation}
In the conventional version of the QVC model the phonon vertices
$\gamma_{\mu}$ are derived from the corresponding transition
densities $\mathcal{R}_{\mu}$ and the static effective interaction:
\begin{equation}
\gamma_{\mu;k_{1}k_{2}}^{\eta_{1}\eta_{2}}=\sum\limits_{k_{3}k_{4}}%
\sum\limits_{\eta_{3}\eta_{4}}\tilde{V}_{k_{1}k_{4},k_{2}k_{3}}^{\eta_{1}%
\eta_{4},\eta_{2}\eta_{3}}\mathcal{R}_{\mu;k_{3}k_{4}}^{\eta_{3}\eta_{4}},
\label{phonon}%
\end{equation}
where
$\tilde{V}_{k_{1}k_{4},k_{2}k_{3}}^{\eta_{1}\eta_{4},\eta_{2}\eta_{3}}$
is defined in Eq. (\ref{static-interaction}).
%
\section{QVC in nuclear response function: relativistic quasiparticle time blocking
approximation}

A response of a superfluid nucleus to a weak external field is
conventionally described by the Bethe-Salpeter equation (BSE)
\cite{SB.51}. The method to derive the BSE for superfluid
non-relativistic systems from a generating functional is known and
can be found, e.g.,
in Ref. \cite{Tse.07} where the generalized Green's function
formalism was used. Applying the same technique in the relativistic
case, one obtains a similar ansatz for the BSE. For our purposes, it
is convenient to work in the time representation: let us, therefore,
include the time variable and the variable $\eta$ defined in Eq.
(\ref{hb}), which distinguishes components in the doubled
quasiparticle space, into the single-quasiparticle indices using
$1=\{k_{1},\eta_{1},t_{1}\}$. In this notation the BSE for the
response function $R$ reads:
\begin{equation}
R(14,23)=G(1,3)G(4,2)-i\sum\limits_{5678}G(1,5)G(6,2)V(58,67)R(74,83),
\label{bse0}%
\end{equation}
where the summation over the number indices $1$, $2,\dots$ implies
integration over the respective time variables. The function $G$ is
the exact single-quasiparticle Green's function, and $V$ is the
amplitude of the effective interaction irreducible in the
$ph$-channel. This amplitude is determined as a variational
derivative of the full self-energy $\Sigma$ with respect to the
exact single-quasiparticle Green's function:
\begin{equation}
V(14,23)=i\frac{\delta\Sigma(4,3)}{\delta G(2,1)}. \label{uampl}%
\end{equation}
Here we introduce the free response $R^{0}(14,23)=G(1,3)G(4,2)$ and
formulate the Bethe-Salpeter equation (\ref{bse0}) in a shorthand
notation, omitting the number indices:
\begin{equation}
R=R^{0}-iR^{0}VR. \label{bse0s}%
\end{equation}
For the sake of simplicity, we will use this shorthand notation in
the following discussion. Since the self-energy in Eq.~(\ref{Sigma})
has two parts ${\Sigma}={\tilde{\Sigma}}+{\Sigma}^{(e)}$, the
effective interaction ${V}$ in Eq.~(\ref{bse0}) is a sum of the
static RMF interaction ${\tilde{V}}$ and the energy-dependent term
${V}^{(e)}$:
\begin{equation}
V={\tilde{V}}+{V}^{(e)},
\end{equation}
where (with $t_{12}=t_{1}-t_{2}$)
\begin{equation}
\tilde{V}(14,23)=\tilde{V}_{k_{1}k_{4},k_{2}k_{3}}^{\eta_{1}\eta_{4},\eta
_{2}\eta_{3}}\delta(t_{31})\delta(t_{21})\delta(t_{34})\,, \label{V-static}%
\end{equation}%
\begin{equation}
V^{(e)}(14,23)=i\frac{\delta\Sigma^{(e)}(4,3)}{\delta G(2,1)}\,,
\label{e-interaction}%
\end{equation}
and\textsl{ $\tilde{V}_{k_{1}k_{4},k_{2}k_{3}}^{\eta_{1}\eta_{4},\eta_{2}%
\eta_{3}}$ }is determined by
Eq\textsl{.~}(\ref{static-interaction})\textsl{. }
In the DHBCS basis the Fourier transform of the amplitude $V^{(e)}$
has the form:
\begin{equation}
V_{k_{1}k_{4},k_{2}k_{3}}^{(e)\eta_{1}\eta_{4},\eta_{2}\eta_{3}}%
(\omega,\varepsilon,\varepsilon^{\prime})=\sum\limits_{\mu,\eta_{\mu}}%
\frac{\eta_{\mu}\gamma_{\mu;k_{3}k_{1}}^{\eta_{\mu};\eta_{3}\eta_{1}}%
\gamma_{\mu;k_{4}k_{2}}^{\eta_{\mu};\eta_{4}\eta_{2}\ast}}{\varepsilon
-\varepsilon^{\prime}+\eta_{\mu}(\Omega_{\mu}-i\delta)}\ ,\ \ \ \ \
\ \delta
\rightarrow+0\ . \label{ueampl}%
\end{equation}
In order to make the BSE (\ref{bse0s}) more convenient for the
further analysis we eliminate the exact Green's function $G$ and
rewrite it in terms of the mean field Green's function $\tilde{G}$
which is diagonal in the DHBCS basis. In time representation we have
the following ansatz for $\tilde G$:
\begin{equation}
{\tilde{G}}(1,2)=-i\eta_{1}\delta_{k_{1}k_{2}}\delta_{\eta_1\eta_2}\theta(\eta_{1}\tau
)e^{-i\eta_{1}E_{k_{1}}\tau},\ \ \ \ \ \tau=t_{1}-t_{2}.
\label{mfgft}
\end{equation}

Using the connection between the mean field GF $\tilde{G}$ and the
exact GF $G$ in the Nambu form
\begin{equation}
{\tilde{G}}^{-1}(1,2)=G^{-1}(1,2)+\Sigma^{(e)}(1,2),
\end{equation}
one can eliminate the unknown exact GF $G$ from the
Eq.~(\ref{bse0s}) and rewrite it as follows:
\begin{equation}
R={\tilde{R}}^{0}-i\tilde{R}^{0}WR, \ \ \ \ \ \ W=\tilde{V}+W^{(e)}, \label{bse}%
\end{equation}
with the mean-field response ${\tilde{R}}^{0}(14,23)={\tilde{G}}%
(1,3){\tilde{G}}(4,2)$ and $W$ as a new interaction, where
\bea
W^{(e)}(14,23)=V^{(e)}(14,23)+i\Sigma^{(e)}(1,3){\tilde{G}}^{-1}%
(4,2) + \nonumber\\
+ i{\tilde{G}}^{-1}(1,3)\Sigma^{(e)}(4,2)- i\Sigma^{(e)}(1,3)\Sigma^{(e)}(4,2). \label{wampl1}%
\eea
Thus, we have obtained the BSE in terms of the mean-field
propagator, containing the well-known mean-field Green's functions
${\tilde{G}}$, and a rather complicated effective interaction $W$ of
Eqs.~(\ref{bse},\ref{wampl1}), which is also expressed through the
mean-field Green's functions.

Then, we apply the quasiparticle time blocking approximation (QTBA)
to the Eq. (\ref{bse}) employing the time projection operator in the
integral part of this equation \cite{Tse.07}.
The time projection leads, after some algebra and the transformation
to the energy domain, to an algebraic equation for the response
function. For the $ph$-components of the response function it has
the form:
\bea
R_{k_{1}k_{4},k_{2}k_{3}}^{\eta\eta^{\prime}}(\omega)&=&\tilde{R}_{k_{1}k_{2}%
}^{(0)\eta}(\omega)\delta_{k_{1}k_{3}}\delta_{k_{2}k_{4}}\delta_{\eta
\eta^{\prime}} + \nonumber\\ &+& \tilde{R}_{k_{1}k_{2}}^{(0)\eta}(\omega)\sum\limits_{k_{5}%
k_{6}\eta^{\prime\prime}}{\bar{W}}_{k_{1}k_{6}%
,k_{2}k_{5}}^{\eta\eta^{\prime\prime}}(\omega)R_{k_{5}k_{4},k_{6}k_{3}}%
^{\eta^{\prime\prime}\eta^{\prime}}(\omega), \label{respdir}%
\eea
where we denote $ph$-components as
$R_{k_{1}k_{4},k_{2}k_{3}}^{\eta\eta^{\prime}}(\omega) =
R_{k_{1}k_{4},k_{2}k_{3}}^{\eta,-\eta^{\prime},-\eta,\eta^{\prime}}(\omega),
$ and
%
%
\begin{equation}
{\bar{W}}_{k_{1}k_{4},k_{2}k_{3}}^{\eta\eta^{\prime}}(\omega)=\tilde{V}%
_{k_{1}k_{4},k_{2}k_{3}}^{\eta\eta^{\prime}}+\Bigl(\Phi_{k_{1}k_{4},k_{2}%
k_{3}}^{\eta}(\omega)-\Phi_{k_{1}k_{4},k_{2}k_{3}}^{\eta}(0)\Bigr)\delta
_{\eta\eta^{\prime}}. \label{W-omega}%
\end{equation}
In Eq. (\ref{W-omega}) $\Phi (\omega)$ is the dynamical part of the
effective interaction responsible for the QVC with the following
$\eta= \pm1$ components:
\bea%
\Phi_{k_{1}k_{4},k_{2}k_{3}}^{\eta}(\omega)  &=& \nonumber \\
= \sum\limits_{\mu\xi}\delta_{\eta\xi}
\Bigl[\delta_{k_{1}k_{3}}\sum\limits_{k_{6}} \frac{\gamma_{\mu;k_{6}k_{2}%
}^{\eta;-\xi} \gamma_{\mu;k_{6}k_{4}}^{\eta;-\xi\ast}}{\eta\omega-E_{k_{1}}-E_{k_{6}%
}-\Omega_{\mu}} &+&
\delta_{k_{2}k_{4}}\sum\limits_{k_{5}}\frac{\gamma
_{\mu;k_{1}k_{5}}^{\eta;\xi}
\gamma_{\mu;k_{3}k_{5}}^{\eta;\xi\ast}}{\eta\omega-
E_{k_{5}} - E_{k_{2}} - \Omega_{\mu}}\nonumber\\
-\Bigl(\frac{\gamma_{\mu;k_{1}k_{3}}^{\eta;\xi} \gamma_{\mu
;k_{2}k_{4}}^{\eta;-\xi\ast}}{\eta\omega- E_{k_{3}}- E_{k_{2}} -
\Omega_{\mu}} &+&
\frac{\gamma_{\mu;k_{3}k_{1}}^{\eta;\xi\ast}\gamma_{\mu;k_{4}k_{2}}^{\eta;
-\xi}} {\eta\omega- E_{k_{1}} - E_{k_{4}} -
\Omega_{\mu}}\Bigr)\Bigr],
\label{phiphc0}%
\eea where we denote $\gamma_{\mu;k_{1}k_{2}}^{\eta;\xi} =
\gamma_{\mu;k_{1}k_{2}}^{\eta;\xi\xi}$.

%
By construction, the propagator $R(\omega)$ in Eq.~(\ref{respdir})
contains only configurations which are not more complicated than
2q$\otimes$phonon.
In Eq. (\ref{W-omega}) we have included the subtraction of $\Phi(0)$
because of the following reason.
Since the parameters of the density functional and, as a
consequence, the effective interaction $\tilde{V}$ are adjusted to
experimental ground state properties at the energy $\omega=0$, the
part of the QVC interaction, which is already contained in
$\tilde{V}$ and given approximately by $\Phi(0)$, should be
subtracted to avoid double counting of the QVC~\cite{Tse.07}.

Eventually, to describe the observed spectrum of an excited nucleus
in a weak external field $P$ as, for instance, an electromagnetic
field, one needs to calculate the strength function:
\begin{equation}
S(E)=-\frac{1}{2\pi}\lim\limits_{\Delta\rightarrow+0}Im\
\sum\limits_{k_{1}%
k_{2}k_{3}k_{4}}\sum\limits_{\eta\eta^{\prime}}P_{k_{1}k_{2}}^{\eta\ast
}R_{k_{1}k_{4},k_{2}k_{3}}^{\eta\eta^{\prime}}(E+i\Delta)P_{k_{3}k_{4}}%
^{\eta^{\prime}}.
\label{strf}%
\end{equation}
The imaginary part $\Delta$ of the energy variable
has the meaning of an additional artificial width for each
excitation and emulates effectively contributions from
configurations which are not taken into account explicitly in our
approach.
\begin{figure}[ht]
\centerline{\psfig{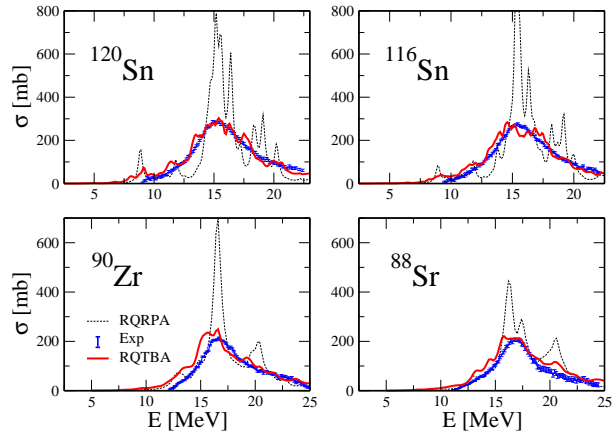}} \caption{Total dipole
photoabsorption cross section in stable medium-mass nuclei, see text
for explanation.}
\label{gdr}
\end{figure}

Fragmentation of the giant dipole resonance (GDR) due to the QVC is
one of the most famous phenomena in nuclear structure physics. To
describe the GDR, one has to calculate the strength function of Eq.
(\ref{strf}) as a response to an electromagnetic dipole operator
which in the long wavelength limit reads:
\be P^{EM}_{1M} =
\frac{N}{A}\sum\limits_{p=1}^{Z}r_pY_{1M}(\Omega_p) -
\frac{Z}{A}\sum\limits_{n=1}^{N}r_nY_{1M}(\Omega_n). \ee
The cross section of the total dipole photoabsorption is given by:
\be \sigma_{E1} = \frac{16\pi^3e^2}{9{\hbar}c}ES(E). \ee
Fig. \ref{gdr} shows the cross sections of the total dipole
photoabsorption in four medium-mass spherical nuclei obtained within
the RQRPA (black dashed curves) and RQTBA (red solid curves),
compared to neutron data (blue error bars) from Ref. \cite{ensdf}.
The details of these calculations are described in Ref.
\cite{LRT.08}. One can clearly see that the QVC included within the
RQTBA provides a sizable fragmentation of the GDR. The QVC mechanism
of the GDR width formation is known for decades, see Refs.
\cite{BBBL.77,Sol.92,KST.04} and references therein. However, the
RQTBA is the first fully self-consistent approach which, in contrast
to the previously developed ones, accurately reproduces the
Lorentzian-like GDR distribution observed in experiments.
\begin{figure}[ht]
\centerline{\psfig{file=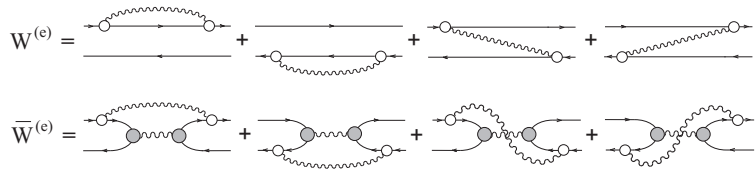,width=10cm}} \caption{The
2q$\otimes$phonon amplitude $W^{(e)}$ of the conventional QVC model
and the two-phonon amplitude $\bar{W}^{(e)}$ of the two-phonon model
in a diagrammatic representation. The solid lines are the
four-component fermion propagators, the wavy curves denote phonon
propagators, the empty circles represent phonon vertices, and the
grey circles together with the two nucleonic lines denote the RQRPA
transition densities.}
\label{twophon}
\end{figure}

The main assumption of the RQTBA discussed so far is that two types
of elementary excitations - two-quasiparticle (2q) and vibrational
modes - are coupled in such a way that configurations of
2q$\otimes$phonon type with low-lying phonons strongly compete with
simple 2q configurations close in energy. There are, however,
additional processes, which are not fully included in this scheme
as, for instance, the coupling of low-lying collective phonons to
multiphonon configurations. Therefore, recently an extension of the
RQTBA has been introduced, which includes also the coupling to
two-phonon states \cite{LRT.10}. In the diagrammatic representation
of the amplitude $W^{(e)}$ of Eq. (\ref{wampl1}) in the upper line
of the Fig. \ref{twophon} the intermediate two-quasiparticle
propagator is represented by the two straight nucleonic lines
between the circles denoting the amplitudes of emission and
absorption of the phonon by a single quasiparticle (the last term of
Eq. (\ref{wampl1}) is omitted because it represents the
'compensating' contribution \cite{KST.04,Tse.07}). In the two-phonon
RQTBA-2 we introduce the RQRPA correlations into the intermediate
two-quasiparticle propagator replacing the amplitude $W^{(e)}$ by
the new one ${\bar W}^{(e)}$.
\begin{figure}[ht]
\vspace{-2cm} \centerline{\psfig{file=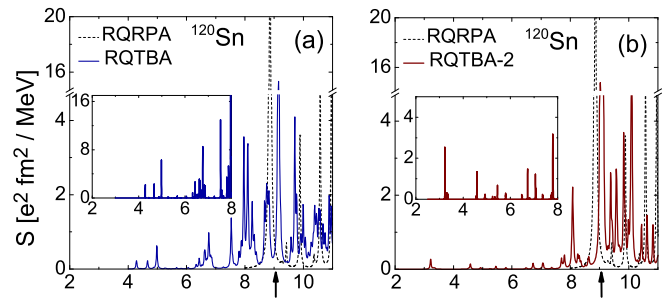,width=10cm}}
\vspace{-0.5cm}\caption{Low-lying dipole spectra of $^{120}$Sn
calculated within the RQRPA (dashed curves), RQTBA (blue solid curve
(a)) and RQTBA-2 (red solid curve (b)). A finite smearing parameter
$\Delta$ = 20 keV has been used in the calculations. The inserts
show the zoomed pictures of the spectra below 8 MeV with a small
value $\Delta$ = 2 keV allowing to see all the states in this energy
region. The arrows indicate the neutron threshold.} \label{sn_tba2}
\end{figure}
Fig. \ref{sn_tba2} illustrates the effect of two-phonon correlations
on spectra of nuclear excitations. It displays the dipole strength
functions for $^{120}$Sn calculated within the conventional RQTBA
and the two-phonon RQTBA-2. The resulting strength functions are
compared with the RQRPA strength function because both of them
originate from the RQRPA by similar fragmentation mechanisms. The
major fraction of the RQRPA state at the neutron threshold (pygmy
mode) shown by the dashed curve is pushed up above the neutron
threshold by the RQTBA-2 correlations. The lowest 1$^-$ state, being
a member of the $[2^+\otimes 3^-]$ quintuplet, appears at 3.23 MeV
with B(E1)$\uparrow$ = 15.9$\times$10$^{-3}$ e$^2$ fm$^2$. These
numbers can be compared with the corresponding data for the lowest
1$^-$ state: it is observed at 3.28 MeV with B(E1)$\uparrow$ =
7.60(51)$\times$10$^{-3}$ e$^2$ fm$^2$, \cite{Bry.99} and
B(E1)$\uparrow$ = 11.20(11)$\times$10$^{-3}$ e$^2$ fm$^2$
\cite{Oze.07}. The obtained agreement with the data is very good in
spite of the fact that these tiny structure at about 3 MeV originate
by the splitting-out from the very strong RQRPA pygmy state located
at the neutron threshold, due to the two-phonon correlations
included consistently without any adjustment procedures. The
physical content of the two-phonon RQTBA reminds the two-phonon
quasiparticle-phonon model \cite{Sol.92}, however, one-to-one
correspondence has not been established. Also, the obtained
differences between the RQTBA and RQTBA-2 results may occur because
of their limitations in terms of the configuration space. Both
2q$\otimes$phonon and phonon$\otimes$phonon configurations are
limited by only four quasiparticles and, perhaps, on the higher
level of the configuration complexity involving six and more
quasiparticles the differences between the coupling schemes will be
less pronounced. This is supposed to be clarified in the future
studies.

%

%
\section{Outlook}
The old concept of the quasiparticle-vibration coupling has been
implemented on a contemporary basis: as self-consistent extensions
of the relativistic QRPA built on the covariant energy density
functional. In these extensions, the QVC and pairing correlations
are taken into account on the equal footing while the CEDF+BCS
approach provides a convenient working basis for the treatment of
the complicated many-body dynamics. Applications to various nuclear
structure phenomena in ordinary and exotic nuclei illustrate that
the self-consistent implementation of many-body correlations beyond
the CEDF theory represents a successful strategy toward a universal
and precise approach for the low-energy nuclear dynamics.


\end{document}